\documentclass{appolb}
\usepackage{epsfig}
\usepackage{color}
\usepackage{cite}
\begin{document}
\pagestyle{plain}
\newcount\eLiNe\eLiNe=\inputlineno\advance\eLiNe by -1
\title {World Financial 2014-2016 Market Bubbles: \\ Oil Negative - US Dollar Positive}
\author{Marcin W\c atorek$^{1}$\footnote{e-mail: marcin.watorek@ifj.edu.pl}, Stanis{\l}aw Dro\.zd\.z$^{1,2}$, Pawe{\l} O\'swi\c ecimka$^{1}$
\address{
$^1$ Complex Systems Theory Department, Institute of Nuclear Physics, Polish Academy of Sciences, PL--31-342 Krak\'ow, Poland\\
$^2$ Faculty of Physics, Mathematics and Computer Science, Cracow University of Technology, PL--31-155 Krak\'ow, Poland}}

\maketitle

\begin{abstract}
 Based on the Log-Periodic Power Law (LPPL) methodology, with the universal preferred scaling factor $\lambda \approx 2$, the negative bubble on the oil market in 2014-2016 has been detected. Over the same period a positive bubble on the so called commodity currencies expressed in terms of the US dollar appears to take place with the oscillation pattern which largely is mirror reflected relative to oil price oscillation pattern. This documents recent strong anti-correlation between the dynamics of the oil price and of the USD. A related forecast made at the time of FENS 2015 conference (beginning of November) turned out to be quite satisfactory. These findings provide also further indication that such a log-periodically accelerating down-trend signals termination of the corresponding decreases.

\end{abstract}

\section{Introduction}

\begin{figure}[ht]
\epsfxsize 13.0cm
\epsffile{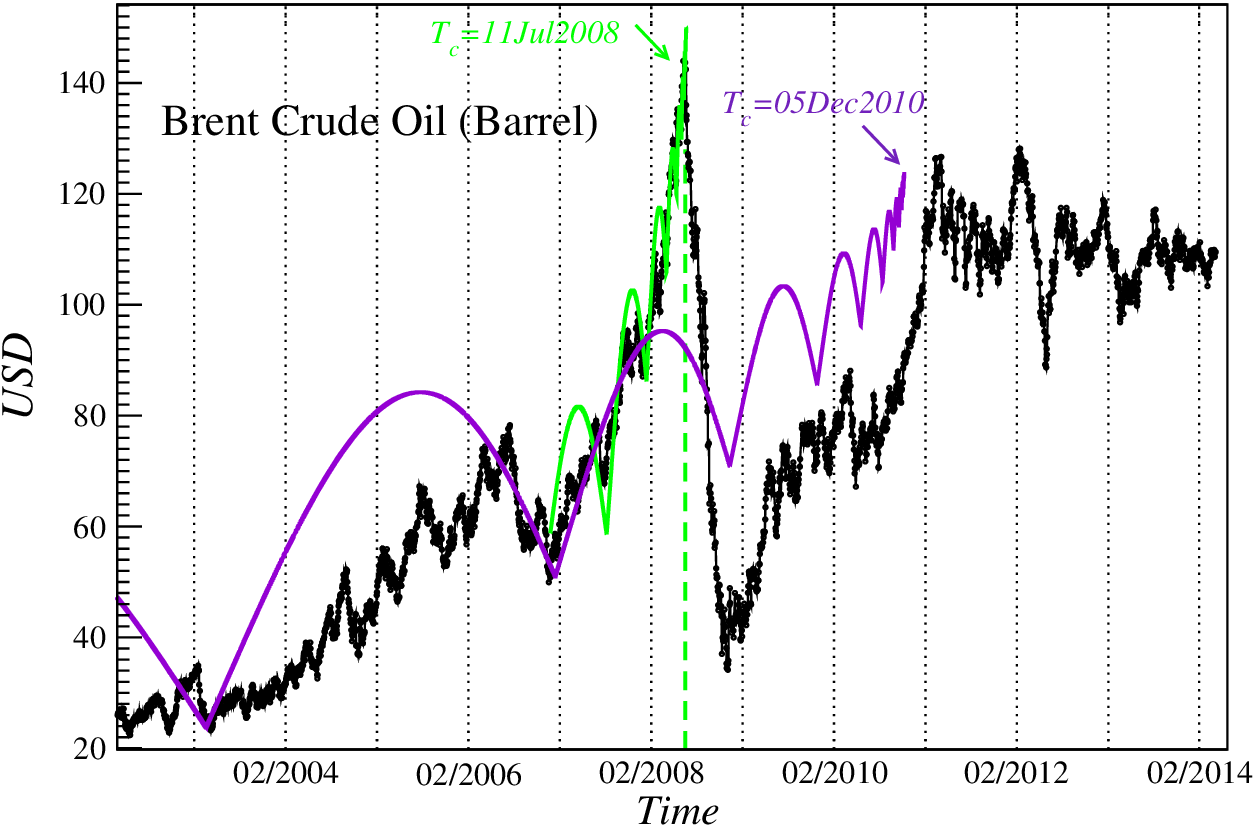}
\caption{Brent Crude Oil log-periodic scenario generated in Spring 2008 \cite{X1} and updated in May 2009 \cite{O1} with actual data from the oil market up to April 2014.}
\label{fig:Brent}
\end{figure}

The concept of financial log-periodicity \cite{Fei96,Sorr96,sornette1998,V98,D1,C7} often termed as Log-Periodic Power-Law (LPPL) model, has widely been used for detecting bubbles and subsequent crashes already for almost two decades. In spite of rising some controversies~\cite{F11, Fei1, Br1}, many successful attempts to describe~\cite{A1,B11,C11,L1,F1,G1,J2,J1,K2,K1,K3,W6,Sorr16} and even to detect bubbles and their subsequent bursts by using this technique~\cite{M1,D21,S1,H1} have been reported. One of the most spectacular such examples is ex-ante exceptionally precise prediction of Brent Crude Oil bubble bursting time in early July 2008, delivered three months ahead as described in ref.~\cite{X1} and also on Wojciech Bia{\l}ek blog \cite{Bi5}. Crucial in this connection was application of the universal preferred scaling factor $\lambda \approx 2$ \cite{D1,C7,B11} and decomposition of the entire oil-price development into long-term trend and a local super-bubble - general concept introduced in ref. \cite{C7} - here operating on the oil price in the first half of 2008 and violently terminating on July 11th, 2008, precisely as predicted. In longer terms the prediction also was that after this super-bubble burst the oil price will return to the longer-term still increasing trend with its ultimate termination in the second half of 2010. A minimally updated variant of the original prediction for this long-term oil development scenario as Figure 5 in ref.~\cite{O1} was presented during FENS 4 conference in May 2009. Exactly this same scenario with the actual oil price course up to the beginning of 2014 is shown in Figure \ref{fig:Brent} of the present contribution. Clearly, there is lot of truth even in this long-term forecast. As predicted, the oil price after recovery from the 2008 super-bubble burst went up sharply until the turn of 2010/2011 and this was the end of this long-term increasing trend, indeed. The following decline was probably at least partly delayed and slowed down by the Arab Spring in the years 2010 - 2013 \cite{AS,AS1,AS2}. The real decrease on the oil market started in mid 2014 and within less than 2 years it dropped by $75\%$ from 106\$ to 26\$ per barrel. Usually such a downward trend is associated with the decelerating log-periodic oscillations but in contrast to most of the previous cases \cite{B11,H2,Z1,Z2,Z3} this phase on the oil market appears to be dominated by the accelerating log-periodic oscillations. Simultaneously and in parallel a positive bubble on the so called commodity currencies expressed in terms of the US dollar (USD), exceptionally strongly anti-correlated with the oil price, has developed. This last period of the oil market dynamics is the main subject of the present contribution.

\section{LPPL model for bubbles}
\label{se1}
The concept of financial log-periodicity is based on the assumption that the financial dynamics is governed by phenomena
analogous to criticality in the statistical physics sense. In its conventional form criticality implies a scale invariance which, for a properly defined function $F(x)$ characterizing the system, means that:
\begin{equation}
F(\lambda x) = \gamma F(x).
\label{eq:F}
\end{equation}
A constant $\gamma$ in this equation reflects how the properties of the system change when it is rescaled by a factor $\lambda$. The general solution of Eq.~(\ref{eq:F}) reads:
\begin{equation}
F(x) = x^{\alpha} P({\ln (x) / \ln(\lambda)}),
\label{eq:logper}
\end{equation}
where the first term represents a standard power-law as it is characteristic
to continuous scale-invariance with the critical exponent  $\alpha = \ln(\gamma)/\ln(\lambda)$ and $P$ denotes a periodic function of period one. This general solution can be interpreted in terms of discrete scale invariance. Due to the second term the continuous dominating scaling acquires a correction that is periodic in $\ln(x)$. It is then meaningful to define $x = \vert t - t_c \vert$, where $t$ denotes the ordinary time labeling the original price time series. This variable $x$  represents a distance to the critical point $t_c$. The resulting spacings between the corresponding consecutive repeatable structures at $x_n$ (i.e., minima or maxima) of the log-periodic oscillations seen in the linear scale follow a geometric contraction according to the relation $\lambda= {x_{n+1} - x_n \over x_{n+2} - x_{n+1}}$. The time points $t_c$ thus correspond to the accumulation of such oscillations and, in the context of the financial dynamics such points indicate a reversal of the trend.
One possible representation of periodic function $P$ is the first term of its Fourier expansion:
\begin{equation}
P(\ln(x)/\ln(\lambda)) = A + B \cos ({\omega \over 2\pi} \ln(x) + \phi).
\label{eq:pfo}
\end{equation}
This implies that $\omega = 2\pi / \ln(\lambda)$\cite{C7}.

\section{Negative bubble}
\label{se2}

One possible mechanism that gives rise to such log-periodic structures is positive feedback. This phenomenon leading to an increasing amplitude of the price momentum can also occur in a downward price regime and, as a result, a faster than exponential downward acceleration can take place. In a positive bubble, the positive feedback results from over optimistic expectations of future returns leading to self fulfilling but transient unsustainable price appreciations. In a negative bubble, the positive feedback reflects the rampant pessimism fueled by short positions leading investors to run away from the market which spirals downwards also in a self fulfilling process. The symmetry between positive and negative bubbles is obvious for currencies. If a currency A strongly appreciates against another currency B following a faster than exponential
trajectory, the value of currency B expressed in currency A will correspondingly fall faster than exponentially in a downward spiral. In this example, the negative bubble is simply obtained by taking the inverse of the price~\cite{N2}.

An alternative related mechanism could be the herding behavior between hedge funds or investors which leads to extreme short positioning building up in the futures market. This regime is unstable and almost anything could trigger short squeeze which leads to rapid price growth. It was precisely this situation that existed in the oil market by the end of 2015~\cite{W1}.

\section{Adjusting procedure}
\label{se3}
In the time domain the Eq. \ref{eq:pfo} can be rewritten as:
\begin{equation}\label{eq:lppl}
p(t)=A+B(t_c-t)^m+C(t_c-t)^m\cos(\omega\ln(t_c-t)-\phi).
\end{equation}

This log-periodic power law (LPPL) model is described by 3 linear parameters ($A, B, C$) and 4 nonlinear parameters ($m, \omega, t_c, \phi$). These parameters are subject to the following constrains as proposed by Sornette \cite{B1}: $0 < m <1$, $6 \leq\omega\leq 13$, $B < 0$, $|C|<1$, $t \le t_c$.

To fit LPPL function Eq.~\ref{eq:lppl} to empirical data we use procedure proposed by Filimonov and Sornette~\cite{C1}, which reduces adjustment to just three nonlinear parameters: $t_c, m, \omega$. The key idea of this method is to decrease the number of nonlinear parameters and simultaneously to eliminate the interdependence between the phase $\phi$ and the angular log-frequency $\omega$.
This one achieves by expanding the cosine term the formula (\ref{eq:lppl}) as follows:
\begin{equation}
\label{eq:lppl_new}
p(t) = A + B(t_c - t)^m + C_1(t_c - t)^m \cos(\omega \ln(t_c -t))+ C_2(t_c - t)^m \sin(\omega \ln(t_c -t)).
\end{equation}

As seen from Eq.\ref{eq:lppl_new}, the LPPL function has now only 3 nonlinear ($t_c, \omega, m$) and 4 linear ($A, B,C_1, C_2$) parameters, and the two new parameters $C_1$ and $C_2$ contain formerly the phase $\phi$.
Based on previous evidence \cite{C7,D1,B11,K3} we are using a constant scaling factor $\lambda \approx 2$, which further reduces the estimation problem ($\omega = 2\pi / \ln(\lambda)$).

In order to fit the LPPL function we select the initial parameters $t_c, m, \omega$. We then calculate linear parameters $A, B, C_1, C_2$ by ordinary least squares method and then minimize the cost function using nonlinear least squares method.
All possible values of start-up parameters: $m \in[0.1, 0.9]$ with step 0.05 and $t_c\in[t+1, t+0.1\*n]$ (where $n$ is the length of time series) with step 5 were tested.
To get more robust results we carried out the analysis on empirical data with moving starting point with the step of 5 trading days in a shrinking time window $[t_1, t_2]$. In our work $t_1$ is changing from 12.06.2014 to 10.07.2014, $t_2$ is fixed on 12.02.2016.
The lowest sum of squared residuals (SSR) points to the best fit within each time window. In fitting process getting a stable value of $t_c$ is essential, therefore we compare the SSR's from each time window by evaluating the mean squared error (MSE). The lowest MSE determines the best fit. In order to further illustrate the stability of the adjusting procedure we present the standard deviation for $t_c$ obtained from all fits with different $t_1$ (std($T_c$) in trading days).
\newpage
\section{Oil versus currency markets}
\label{se4}

 Already a visual chart inspection indicates that in around the end of 2015 the commodity currencies expressed in terms of the US dollar and the oil price  develop similar patterns \cite{W5}. In order to quantify this we calculate the Pearson correlation coefficients from the time series representing the price changes of the currencies and of the Crude Light Oil (CL) in the period June 2014 - March 2016. The results are presented in Table \ref{Corr14}.

\begin{table}[ht]
\centering
\begin{tabular}{|l|l|l|}
\hline
    & CL      & CLdiff  \\ \hline
AUD & -0.9530 & -0.3187 \\ \hline
BRL & -0.8897 & -0.2485 \\ \hline
CAD & -0.9412 & -0.5472 \\ \hline
CLP & -0.9039 & -0.2086 \\ \hline
GBP & -0.9235 & -0.2168 \\ \hline
MXN & -0.9221 & -0.3911 \\ \hline
NOK & -0.9746 & -0.3570 \\ \hline
RUB & -0.9717 & -0.2542 \\ \hline
\end{tabular}
\caption{\label{Corr14} Pearson correlation coefficients of the oil (CL) vs 8 commodity currencies (Australian dollar, Brazilian real, Canadian dollar, Chilean peso, Pound sterling, Mexican peso, Norwegian krone, Russian ruble) in the period 01.06.2014-18.03.2016. 1st column - correlation coefficient calculated from the price time series, 2nd column - correlation coefficient calculated from the corresponding return (CLdiff) time series.
Above results clearly show high correlations between commodity currencies vs USD and oil.}
\end{table}
All these coefficients, even the ones calculated from the returns, are large and negative which reflects the fact that these currencies are anti-correlated with the oil price changes.

\begin{figure}[]
\epsfxsize 13.0cm
\epsffile{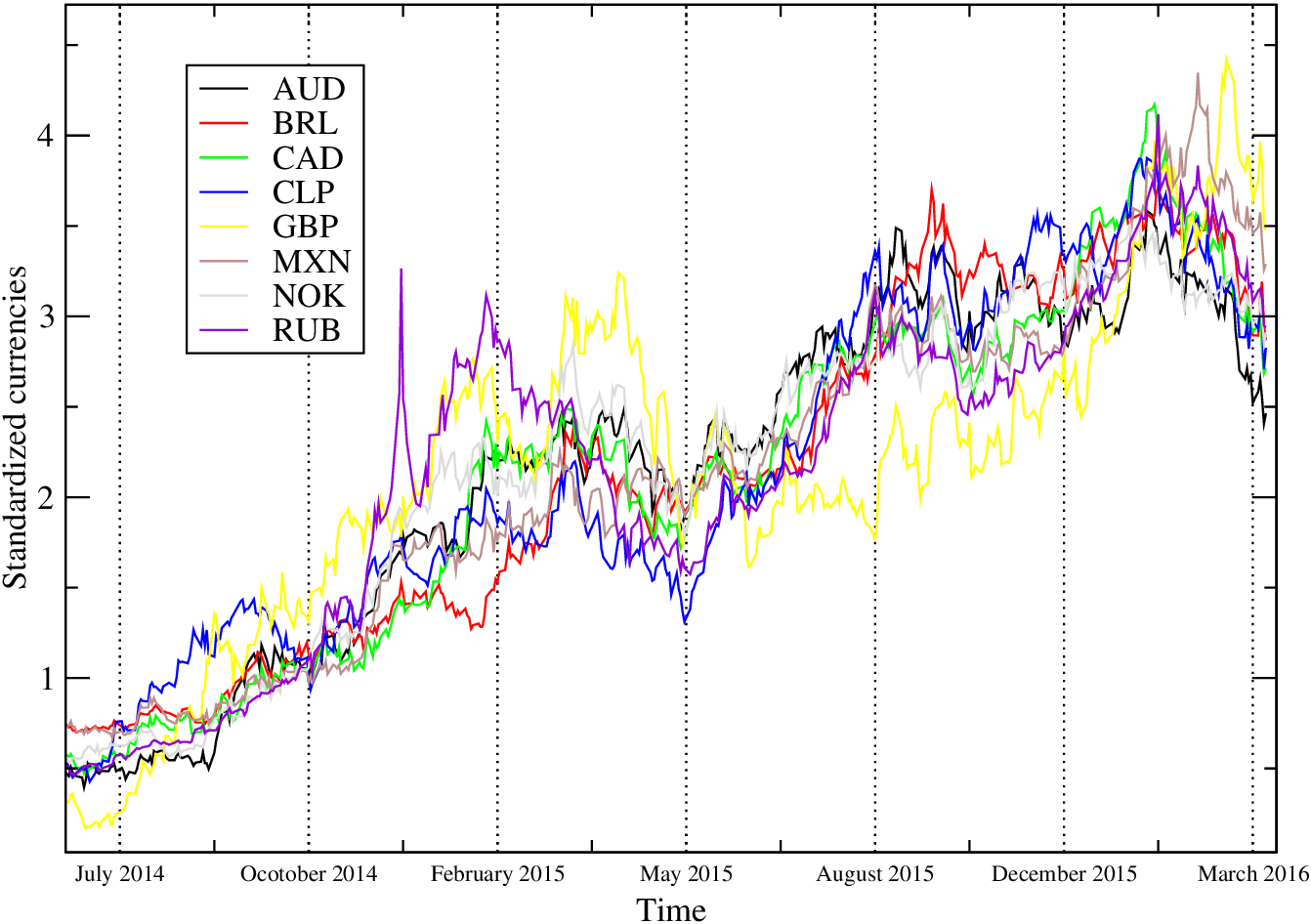}
\caption{Standardized commodity currencies expressed in terms of the US dollar over the period 12.06.2014-18.03.2016.}
\label{fig:Commodtycurr}
\end{figure}

\begin{figure}[ht]
\epsfxsize 13.0cm
\epsffile{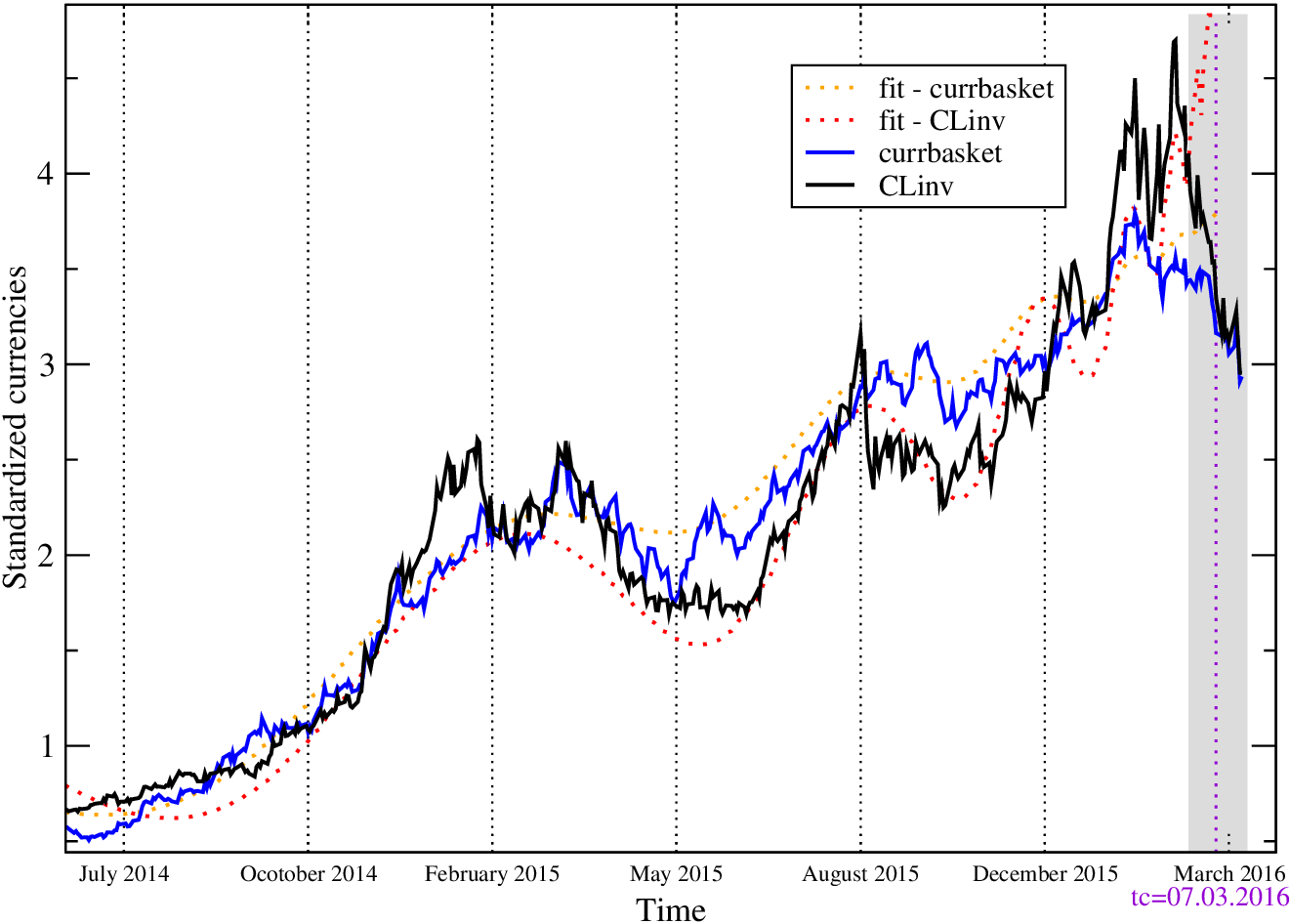}
\caption{Commodity currencies basket standardized (currbasket, blue), the inverse Crude Light Oil price standardized  (CLinv, black) over the period 12.06.2014-18.03.2016 and the corresponding LPPL best fits: fit - currbasket (orange) with the parameters $t_c$=07.03.2016$\pm 11.3$ trading days (std($T_c$) shaded in gray), $m$=0.8926, $\omega$=8.9256, $\lambda$=2.022, MSE=0.0144 and fit - CLinv (red) with the parameters $t_c$=07.03.2016$\pm 9.3$ trading days, $m$=0.2498, $\omega$=8.9317, $\lambda$=2.021, MSE=0.0645. Due to some shifts in phases of the component currencies (Figure \ref{fig:Commodtycurr}) the contracting log-periodic oscillations in the global commodity currencies basket are not as visible as in the single currencies (e.g. Figure \ref{fig:MXNfit}) because of the smoothing effect.}
\label{fig:currbasket}
\end{figure}

 A highly coordinated behaviour of all these currencies expressed in USD can be seen from Figure \ref{fig:Commodtycurr} where they all - in order to make their dynamics directly comparable - are standardized (scaled to have standard deviation 1 and centered to have mean 2). Already visually their oscillatory behaviour quite convincingly follows the same pattern of the log-periodic contractions. For this reason we construct a basket by summing up with equal weight all the considered commodity currencies, i.e. AUD, BRL, CAD, CLP, GBP, MXN, NOK, RUB. The LPPL best fit is performed on this basket and displayed in Figure \ref{fig:currbasket}. The resulting critical time $t_c$=07.03.2016 and as such it was determined already in the beginning of November at the time of FENS 8 Conference. Interestingly, an independent fit performed at the same time to the inverse of the oil price changes, also shown in Figure \ref{fig:currbasket} (standardized in the same way as currencies) points to exactly the same $t_c$. This reflects a highly correlated dynamics of the corresponding time series. This correlation somewhat weakened about five weeks before $t_c$ when the USD reached maximum against the entire basket of all these eight commodity currencies. The inverse oil price reached its highest level some three weeks before this date and started a systematic drawdown. Such a somewhat earlier than $t_c$ burst of the bubble determined by LPPL does not contradict applicability of this methodology and in fact is consistent with the concept of criticality that stays behind LPPL. The closer to $t_c$ is the system the more susceptible it becomes to perturbations that may turn it down~\cite{sornette1998}.

\newpage

\begin{figure}[]
\epsfxsize 13.0cm
\epsffile{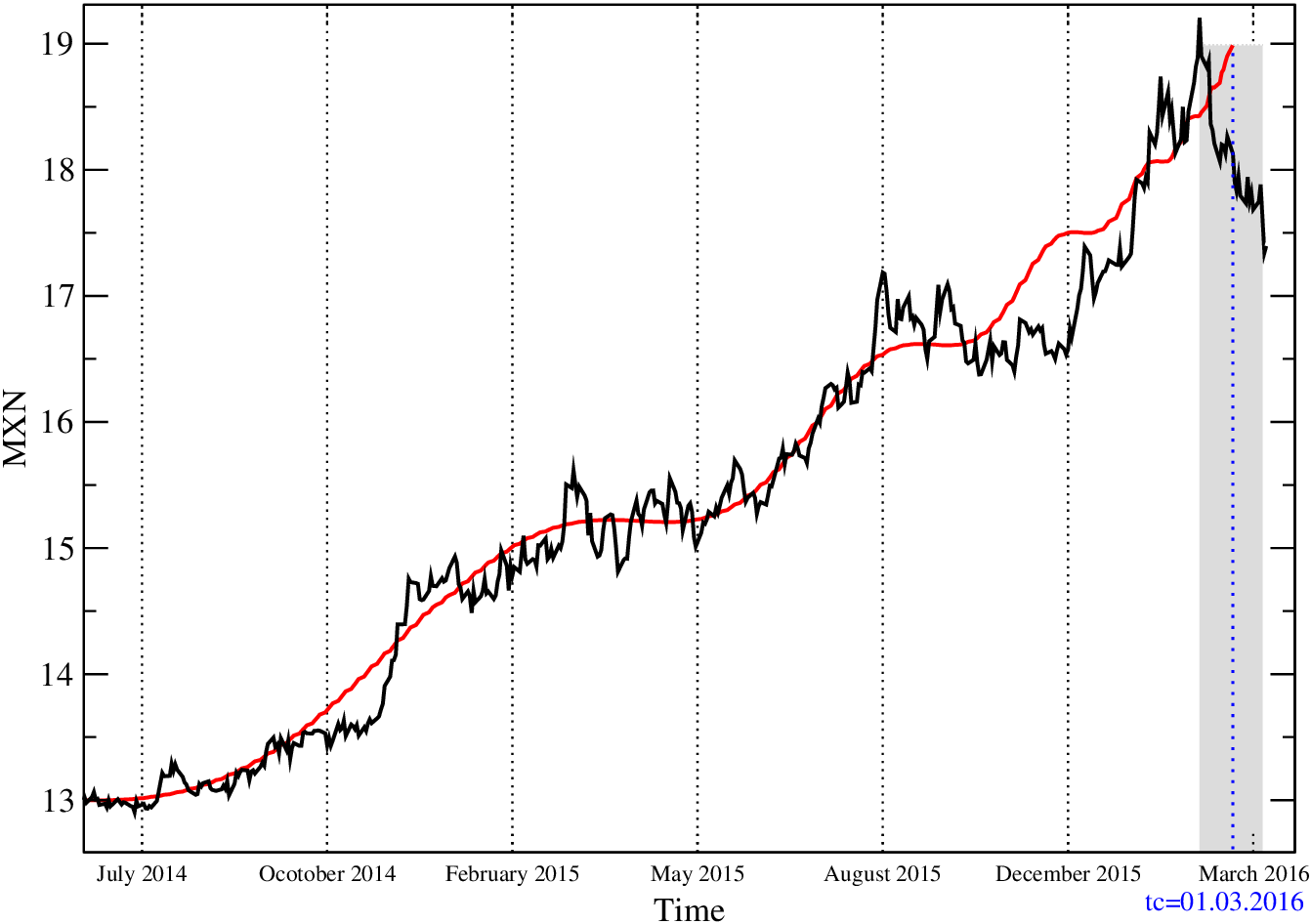}
\caption{Mexican peso vs US dollar in the period 12.06.2014-18.03.2016, best fit (red) parameters are: $t_c$=01.03.2016$\pm 11.8$ trading days (std($T_c$) shaded in gray), $m$=0.6498, $\omega$=8.9964, $\lambda$=2.011, MSE=0.0706. }
\label{fig:MXNfit}

 \end{figure}

\newpage
Not all the currencies in the above commodity basket were equally correlated regarding their way of approaching $t_c$. The highest correlation is observed in the USD expressed in terms of the Mexican peso and for this reason it is shown in a separate Figure \ref{fig:MXNfit}. In this case the trend reversal took place only two weeks before the original prediction.

\begin{figure}[]\
\epsfxsize 13.0cm
\epsffile{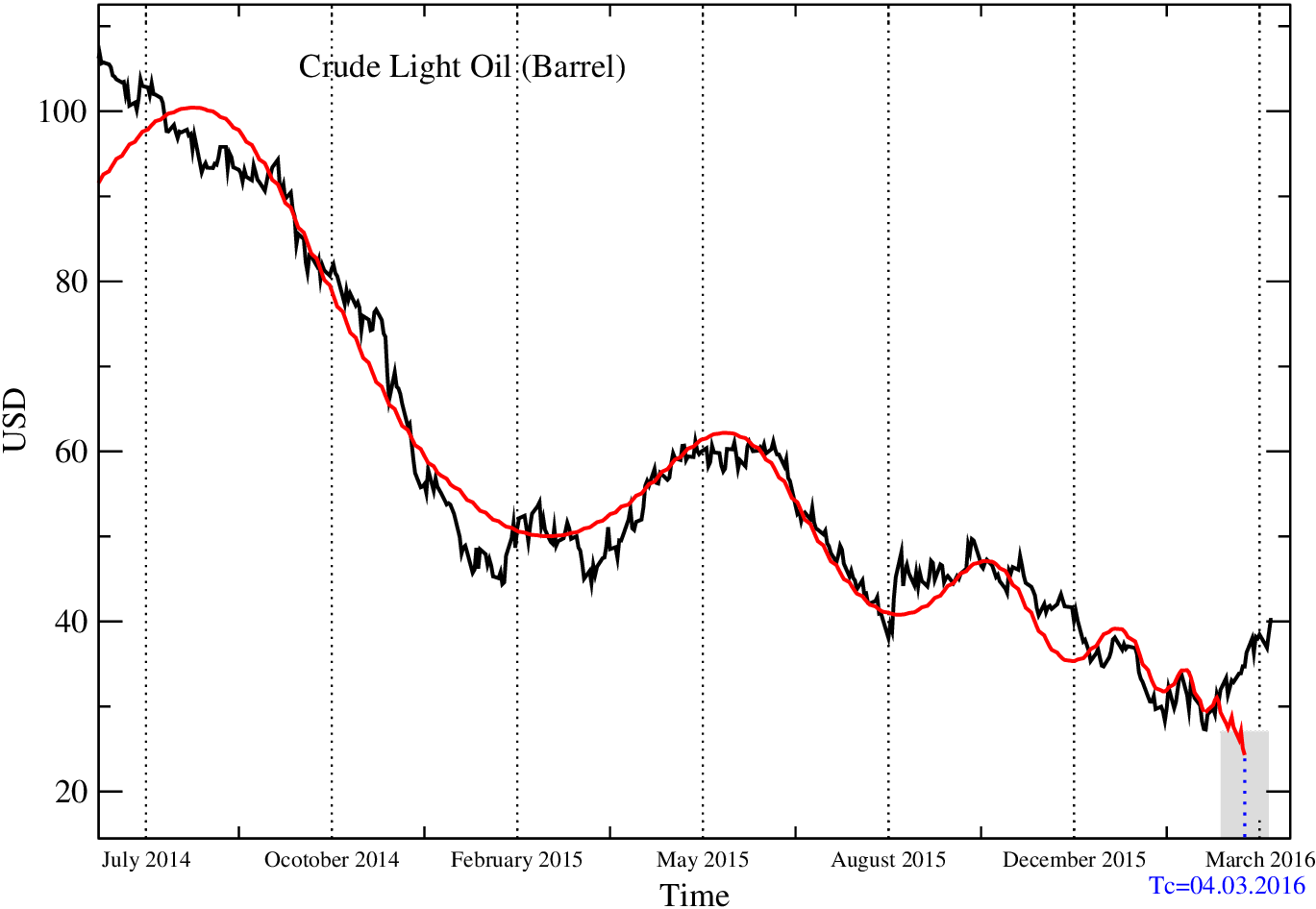}
\caption{Crude Light Oil 12.06.2014-18.03.2016 and its best fit (red) whose parameters read: $t_c$=04.03.2016$\pm 8.8$ trading days (std($T_c$) shaded in gray), $m$=0.2495, $\omega$=8.9817, $\lambda$=2.013, MSE=19.1317. The fit was made in November 2015 at the time of FENS 8 Conference and the oil data updated in March 2016 when the present contribution was under completion.}
\label{fig:CLfit}

 \end{figure}

\newpage
Finally, using the same adjusting procedure as described in section \ref{se3} directly to the Crude Light Oil prices, as displayed in Figure \ref{fig:CLfit}, results in essentially the same critical time $t_c$ as for the inverse oil price and as for the currencies basket. An uncommon feature of this 2014-2016 oil price draw-down is that it is accompanied with the accelerating log-periodic oscillations whose accumulation point signals the real trend reversal which in this case occurred indeed. It therefore belongs to the category of negative bubbles \cite{N2,D5} as confronted with the anti-bubbles \cite{J2,Z1,Z2,Z3}.

\newpage
\section{Summary}
\label{se5}
The downward trend on the world oil market has fully developed starting in mid-2014, thus about four months before the end of quantitative easing in the USA. At around the same time the US dollar started to strengthen. The development of both these markets appears to be describable within the Log-Periodic Power Law methodology with the universal preferred scaling factor $\lambda \approx 2$. A novel aspect of this oil price dynamics is presence of the log-periodically accelerating oscillations accompanying the draw-down phase of the market, therefore termed negative bubbles, contrary to the common scenario where the draw-downs are log-periodically decelerating and are called anti-bubbles. Furthermore, this oil negative bubble appears strongly (anti-)correlated in phase with the US dollar (positive) bubble against the major commodity currencies. Both these bubbles ended in mid-February, 3 weeks before their ultimate limit of termination as set by the critical time $t_c$=07.03.2016. After reaching the low, the Crude Light Oil price surged from 13-year low by $50\%$ in one month. It was the biggest 18-session jump in oil prices over 25 years\cite{bialek}.

\end{document}